\documentclass[preprint,review,12pt,number,sort&compress]{elsarticle}

\usepackage{graphicx}
\usepackage{amssymb}
\usepackage{amsmath}
\usepackage{amsfonts}

\usepackage{url}
\usepackage{hyperref}

\hypersetup{
		colorlinks=true,       		
    	linkcolor=blue,          	
    	citecolor=blue}

\usepackage[nameinlink]{cleveref}

\usepackage[labelfont=bf]{caption}
\usepackage{epsfig}
\begin{document}

\begin{frontmatter}


\title{ Zinc blende ZnS (001) surface structure investigated by XPS,LEED and DFT}



\author[1]{Pablo Oliveira}
\author[2]{Corinne Arrouvel}
\author[1]{Fernando Stavale}

\address[1]{Centro Brasileiro de Pesquisas Físicas, 22290-180, Rio de Janeiro, RJ, Brazil}

\address[2]{DFQM/CCTS, Universidade Federal de São Carlos, Campus Sorocaba, 18052-780 Sorocaba - SP, Brazil}

\begin{abstract}
The formation and stability of a sulfur-rich surface of the zinc blende ZnS (001) single crystal have been examined by x-ray photoemission spectroscopy (XPS), low-energy electron diffraction (LEED), and density-functional theory (DFT) calculations. LEED patterns obtained from ZnS (001) surface prepared in ultrahigh vacuum conditions are compatible to the formation of the ($1 \times 2$) reconstruction. Further XPS investigation is consistent with the conversion of the surface region into a sulfur-rich surface. DFT calculations support the S-richer ZnS surface hypothesis and the indirect notion of increasing in the conductivity of such a new structure 
\vspace{0.25cm}
\end{abstract}


\end{frontmatter}


\newpage
\section{Introduction}
ZnS is a semiconductor of group II-VI with a direct band gap of 3.5-3.9 eV. It is one of the most investigated semiconductors due to its potential of applications in solar energy conversion, photocatalysis, and optoelectronic devices, this later particularly related to bio-sensing and high-frequency electronics \cite{bernard1987electronic,moon2022explosive,asenjo2008study,hoffmann1995environmental,kim2003zinc,cocoletzi2013first}. ZnS may crystallize in two main structures, zinc blende (ZB), known as well as spharelite,  and wurtzite (WZ), with their stability depending on the temperature conditions \cite{shahi2018accurate,grunwald2012transferable}. Its low conductivity and wide band gap are some points that retarding the usage of ZnS in the industry at a large scale, turning the attention of the scientific community towards new mechanisms that solve these questions \cite{blount1966ohmic,woods2020wide} . In this sense, defects engineering is one of the main topics widely explored regarded to ZnS structures from the theoretical point of view. It has been shown that depending on the degree of impurities, at a bulk level, either intrisic or extrinsic doping can alter the semiconducting nature (n-type or p-type), reduces the gap, and increase the range of light absorption response.\cite{zhang2001intrinsic,xiao2014magnetism,liu2020understanding,hoang2019defect,huang2018enhancement,kurnia2015band,sarkar2009enhanced,d2017new,karar2004structure}

Of particular interest in this work, ZnS properties at a surface level are also doping-scalable. For instance, water affinity, interaction and stability of amino acids, and reactivity properties as $CO_{2}$ photoreduction can be drastically modified depending on the nature and concentration of the impurities \cite{meng2013dependence,li2019dft,sahraei2020surface,hamad2005simulation,long2016adsorption,sahraei2021chemical,pang2019cation}. Another possibility is geometrical modifications on the surface depending on cation or anion-rich scenario. In particular, polar surfaces (Tasker type III) like ZnS surfaces are expected to undergo crystallographic reconstruction according to the degree of defects (cationic or anionic) on the surface \cite{tasker1979stability}. This issue is widely reported to oxide metal surfaces \cite{noguera2000polar} and predicted from force field simulations for cation-defectiveness (anion-rich) ZnS (001) and (111) polar surfaces \cite{wright1998simulation}, but there is a lack of experimental realization that certifies such an expectation for ZnS system in the most stable phase, and that can explain the driven-force behind such modifications.


The understanding on ZnS surface properties from the experimental point of view is limited due to the mentioned poor conductivity character of these systems, which limits the employment of traditional surface investigations based on XPS and LEED analysis \cite{pesty2005low}. To overcome this problem, thin-film growth on a metallic substrate at different pressures and temperatures is a mechanism largely explored
\cite{deng2017single,zakerian2018investigation,patel2018thermal,yang2019effects,shan2019effect,lonkar2018facile,zhang2003molecular}.
The thickness of the film can be controlled by adjusting the sample preparation scenario, like the annealing temperature that seems to play a role on the electronic response of the sample at reduced dimensionality by increasing the major carriers (electrons or holes) and decreasing the band gap through the formation of defects on the surface \cite{krainara2011structural,pankratov1993hubbard,spicer1976synchrotron,bennett1981si}. However, the dimensionality is another parameter that can alter the arrangement of the sample, since depending on the thickness of the film a high-level of film-substrate interaction is obtained, which might driving geometrical modifications on the sample like a surface reconstruction \cite{deng2017single}. Therefore, it is hard to disclose unequivocally the pathway to modify ZnS surface properties through that alternative. Instead, the mechanism behind the influence of thermodynamic conditions and point defects on ZnS surface would be individually disclosed by working with a high-purity single crystal and relating experimental results with theoretical calculations. Nevertheless, efforts in this direction were only applied to WZ-ZnS \cite{pang2019cation,deng2017single} systems and have not been explored yet for ZB ZnS surfaces.



Herein, we investigate the surface structure of ZnS (001) single crystals by XPS, LEED, and DFT calculations. This research aims to disclose the role played by preparation

conditions on the structural and electronic properties of that system. Our findings reveal an S-rich - reconstructed surface characterized by missing rows of zinc cations. Planar sulfur anions are located notably in the topmost layers due to bonding configurations that drive the surface to a local surface energy minimum consisting of low zinc and high sulfur chemical potential, which seem to be the best configuration in terms of surface conductivity. These results are addressed to the annealing temperature in which the ZnS single crystal was prepared

\section{Methodology}
\label{S:2}
\subsection{Experimental and Computational methods}

The experiments were carried out in an ultrahigh vacuum (UHV) system equipped with standard surface preparation facilities and XPS (SPECS PHOIBOS 150) and LEED. The base pressure was maintained at $\approx 5 \times 10^{-10} \ mbar $. The pass energy used were $ 50 \ eV $, and $ 20 \ eV $, for survey and high-resolution measurements, respectively, with a combined analyzer resolution for the high-resolution spectra given by $0.7 \ eV$, obtained using a monochromatic $Al- K\alpha $ source. Some XPS spectra were collected with flood gun assistance setting the beam energy and the emission current to $1.5 \ eV$ and $18  \ \mu A$, respectively. ZnS (001) single crystals (Mateck GmbH) were prepared after several cycles of ion sputtering (at $600 \ eV$) and annealing from $500$ up to $1420 \ K$. LEED measurements were performed in the preparation chamber before the collection of the photoemission spectra (PE), and the setup was calibrated with a $Ar^{+}$ clean Au (111) single crystal taking as reference to the PE spectra the  $Au 4f_{7/2}$ binding energy at $84 \ eV$ . All the experimental results were obtained at room temperature.

\subsection{Computational Method}

First-principles calculations were carried out to investigate the stability of ZnS surface with and without reconstruction. The structure has been fully optimized (relaxation of the ions and cell parameters) within DFT formalism implemented in the Viena Ab-initio Simulation Package (VASP) software using a   $12 \times 12 \times 12$ gamma-centered k-points grid with RPBE+D3 functional and Projector Augmented Wave (PAW)  pseudopotential \cite{kresse1999ultrasoft}. The cutoff on the kinetic energy is $500 \ eV$. The convergence criterion for the self-consistent electronic cycle is fixed at $10^{-6} \ eV $  per cell. The optimization of atomic geometry at $0 \ K$ is performed by determining the exact Hellman-Feynman forces that act on the ions for each optimization step and using a conjugate gradient algorithm. The convergence criterion for the geometric cycle is $10^{-4} \ \ eV$ . On the $1 \times 1$ surface slab, all ions are fully relaxed using the k-points grid of $12 \times 12 \times 1$ (the volume and the shape of the cell was kept fixed). The slab thickness before relaxation is 14.8 \AA \ and the vacuum thickness of 13.2 \AA. The $2 \times 2$ surface supercell (with $40$ ZnS units) has been simulated with the $6 \times 6 \times 1$ k-points grid. The surface energy $\gamma$, of all structures were calculated as follows :
\begin{equation}
    \gamma_{hkl} = \frac{E_{hkl} - NE_{bulk}}{2A_{hkl}} ,
    \label{surf-energy}
\end{equation}
where $E_{hkl}$ is the energy of the $(hkl)$ slab, $E_{bulk}$ the bulk energy normalized to the number $N$ of ZnS units in the supercell, and $A_{hkl}$ the surface area.  In this study, the (001) ZB-ZnS surface was investigated, then it will be denoted as $\gamma_{001} $. XPS analysis is done with the calculation of binding energies in (Z+1) and Janak approximation, both implemented in VASP software. The excitation state energy of the $S (2p)$ surface species at the final state referred to S bulk (in the middle of the slab). For a better energetic criterion, we have a convergence of $10^{-5} \ eV/cell$ for the electronic and $0.02 \ eV/$\AA \ on forces for the geometry. 

\section{Results and Discussion}

\subsection{Identifying surface chemical species }
\begin{figure}[ht!]
\centering
\includegraphics[scale=0.52]{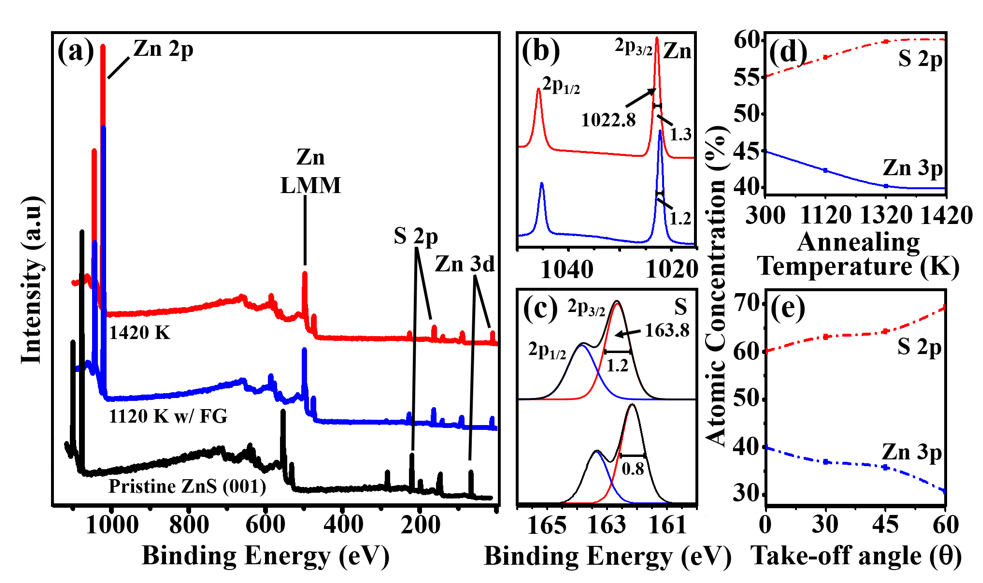}
\caption[Figure 1-]{(a) XPS survey spectra of clean ZnS (001) pristine, ZnS (001) with flood gun and ZnS (001) after a high surface treatment with annealing temperature at 1420 K (black, blue and red curves,respectively). (b) Zn 2p and (c) S 2p core-level spectra.  Atomic concentration as a function of (d) Annealing Temperature and (e) Take-off angle. The spectra are offset for clarity. All the spectras were acquired at room temperature.}
\label{xps}
\end{figure}

 Several XPS spectra obtained of the ZnS (001) single crystal are depicted in \cref{xps}. First, in \cref{xps}(a), are shown some spectra of three differents scenarios. Initially, the sample was measured as-loaded . In this scenario, the main core level peaks are located at $1074 \ eV$, $678\  eV$ and $203 \ eV$, corresponding to Zn $2p$, Zn $LMM$ and S $2p$, respectively. These binding energies are significantly higher than their typical values previously reported \cite{dengo2020depth}. This fact might be addressed to the insulator nature of ZnS, which implies a low electron-hole recombination rate, resulting a strong electrostatic interaction between the ejected photoelectron and the positive charge accumulated at the surface. This interaction is responsible to the decrease of the photoelectron kinetic energies thereby leading the peaks towards higher binding energies positions.

 
 When several cycles of sputtering and annealing were conducted at an annealing temperature of $1120 \ K$, it was noted an appreciable reduction in that shift. The binding energy of the Zn $2p$ component arises at $1031.8 \ eV$, whereas the Zn $LMM$ and S $2p$ ones at $510$ and $173 \ eV$, correspondingly, then configuring a shift at about $10 \ eV$, a smaller value than that reported by Barreca \cite{barreca2002analysis}. With flood gun assistance, the remaining charges were compensated and the main core-level peaks have arisen with their characteristic binding energies: Zn $2p$ at $1022 \ eV$, Zn $LMM$ and $S \ 2p$ at $500$ and $163.1 \ eV$ respectively. A Further increase at the sample annealing temperature from $1120 \ K$ up to $1620 \ K$, reveals XPS spectra quite close to the expected ZnS core-level peak positions, even without flood gun assistance, as highlighted by the equivalence of both spectras. This finding is suggestive for an improvement on the conductivity at the ZnS surface.
Detailed analysis of these distinguished situations are obtained from the high resolution (H.R) spectra of the Zn $2p$ and S $2p$ components, depicted in \cref{xps}(b) and \cref{xps}(c), respectively ( Zn LMM and Valence Band are shown in the supp. material). In \cref{xps}(b) are shown two characteristic peaks associated to Zn $2p_{1/2}$,and Zn $2p_{3/2}$ components. With flood gun assistance (blue line spectrum) these peaks lie at $1045.2$ and $1022.2 \ eV$, respectively, representing a doublet separation of $ 23 \ eV$, a typical value for Zn ions binding to sulfur anions in ZnS structure \cite{barreca2002analysis,liang2018surface}. Post high temperature sample annealing treatments (red line spectrum), the Zn main peaks lie almost with the same binding energy: $1022.8 \ eV$ and $1045.8 \ eV$ for Zn $2p_{3/2}$ and Zn $2p_{1/2}$ components, respectively, preserving the spin-orbit split. Similar findings are observed from  S $2p$ H.R spectra presented in \cref{xps}(c).  The spectrum acquired in presence of ion gun (bottom one) reveals the S 2p components at $163.4 \ eV$ and $162.2 \ eV$, corresponding to S $2p_{1/2}$ and S $2p_{3/2}$, respectively, indicating sulfur atoms at $S^{2-}$ state in ZnS lattice, as discussed in previous work \cite{dengo2020depth}. The spin-orbit split and $1:2$ ratio of such peaks are unchanging under high surface treatment, as shown in the top spectra. The S $2p_{3/2}$ component lies at $163.8 \ eV$, $1.2 \ eV$ away from the S $2p_{1/2}$ one, which agrees very well with the previous experiment. Furthermore, the $FHWM$ depicted in both figures are in good agreement with literature results of sulfide structures \cite{laajalehto1994xps}. Interestingly, the broadening of S 2p peak suggest sulfur anions segregating at the surface due to the removal of Zn species post the high surface treatment. Such a hypothesis is inspected from an specific analysis of the atomic concentration at the ZnS surface as a function of take-off angle and annealing temperature, as present in \cref{xps}(d) and \cref{xps}(e), respectively. The quantification showed in both figures is done by comparing Zn $3p$ and S $2p$ peaks, because the photoelectrons of these components are ejected at approximately the same depth sample. From \cref{xps}(d) one can inspect that  higher annealing temperature the higher $S \ 2p \ / Zn \ 3p$ ratio. From the initial process to the higher treatments scenario, an increase of $5\%$ in S $2p$ concentration is observed, then indicating the preparation condition drives sulfur–enrichment at the ZnS surface. The take-off geometry increases the surface sensibility, therefore the dominancy of sulfur over zinc depicted in \cref{xps}(e) is explained in terms of an even S-richer scenario at the outermost layers.

\newpage
\subsection{Geometrical modifications and theoretical results}

\begin{figure}[htb!]
    \centering
    \includegraphics[scale=0.6]{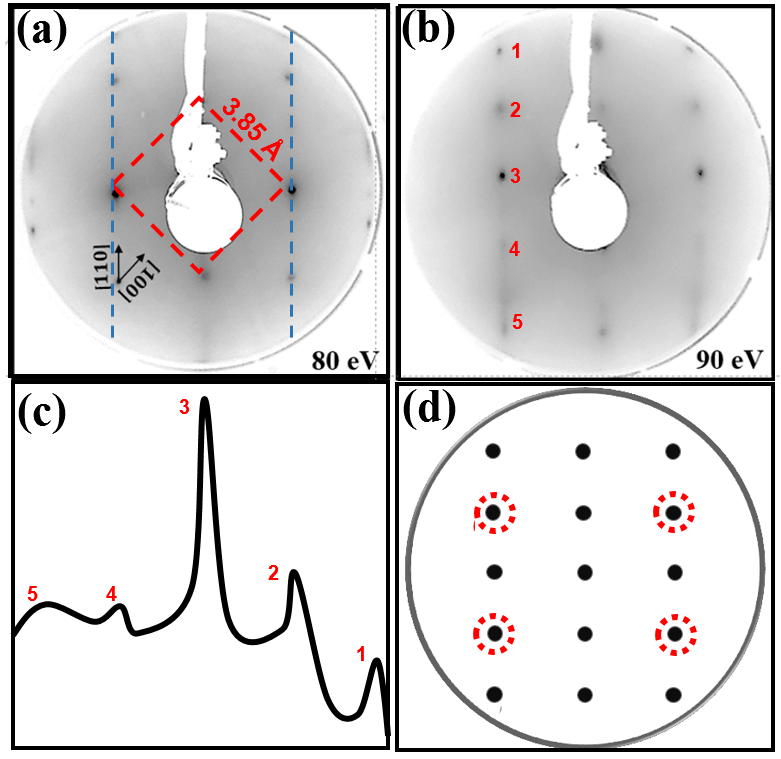}
    \caption{LEED patterns of ZnS (001) single crystal at (a) 80 eV and (b) 90 eV ;(c) line profile plot from the enumerated points in (b) ; (d) LEEDpat simulation of a (1 x 2) reconstruction.  All LEED patterns were adjusted to enhance the image contrast}
    \label{fig:LEED}
\end{figure}

\newpage

Additional information about the ZnS surface structure is obtained from LEED measurements and analysis displayed in \cref{fig:LEED}. From \cref{fig:LEED}-(a) one can observe the Bragg spots are arranged in a tetragonal symmetry resulting in a square reciprocal unit cell. The lattice parameter of the surface unit cell is $3.85$ \AA \  \footnote{See supp.Material to understand how the lattice parameter is estimated}, in good agreement with theoretical results and databases. In addition, one can note high directional rows of points along [110] direction (blue-dashed line) whose length is almost twice the surface unit cell length, thereby indicating a translational symmetry breaking and thus suggesting a $(1 \times 2)$ reconstruction of the ZnS surface. This result is ratified in \cref{fig:LEED}-(b) in which two extra Bragg spots (labeled '2' and '4', respectively) along the [110] direction are present. The five experimental points along [110] are highlighted from a profile plot showed in From \cref{fig:LEED}-(c). The asymmetry on the intensity of the Bragg spots is explained by the charging at the ZnS surface that turns some spots to lie blurry at some energies and that hampered the achievement of a well-defined LEED pattern before the high-temperature surface treatment (supp.material). The experiment-suggested $(1 \times 2)$ surface reconstruction is reproduced through the LEEDpat simulation \cite{hermannleedpat} and shown in figure 2-(c), confirming the shape of the structure with five spots in each row emphasized by the extra red-circled two points. 

\begin{figure}[ht!]
    \centering
    \includegraphics[scale=0.44]{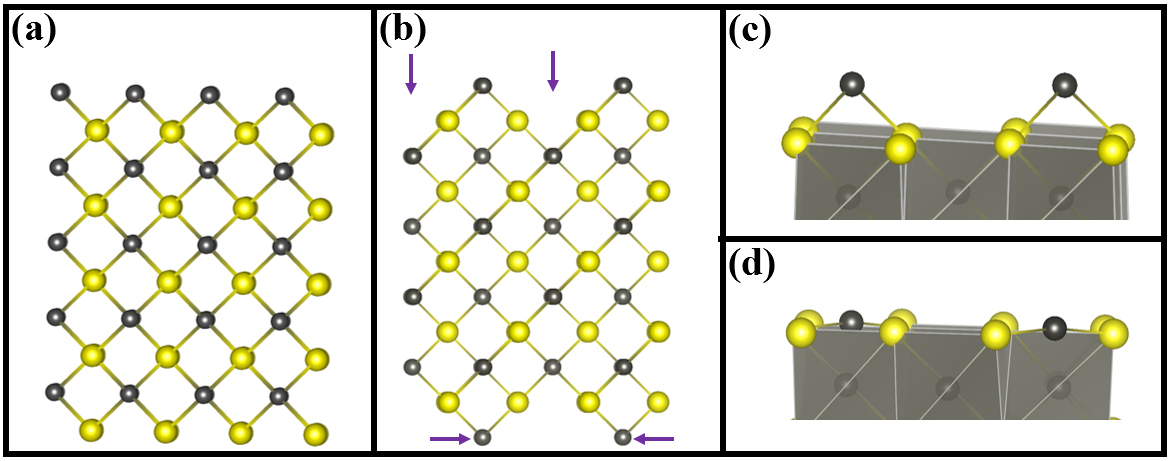}
    \caption{Side views of (a) unreconstructed and (b) reconstructed ZnS (001) surface. Purple arrows are indicating the modifications on the structure ;  Zn-terminated ZnS reconstructed surface (c) before and (d) after ions relaxations. Dark gray and yellow balls are denoting zinc and sulfur, respectively}
    \label{fig:models}
\end{figure}

In this scenario, it is expected Zn and S atoms relaxing inward and outward to the surface, respectively, a typical result derived by geometrical reconstruction of Tasker type III surfaces like ZnS \cite{wright2004interatomic}. Furthermore, this situation increases the number of dangling bonds at the surface, driving surface states that might tail the gap. These findings are corroborative to the previous discussion of the XPS results that suggested a S-rich (especially on the topmost layers) and more semiconducting ZnS surface. To explore these results from the theoretical point of view, an unreconstructed ZnS (001) surface (\cref{fig:models}-(a)), and a reconstructed supercell (\cref{fig:models}(b)) are investigated from first-principles calculations.

\newpage
The unreconstructed $ (1 \times 1) $ slab has an average surface energy $\gamma_{001}$ of $1997 \ mJ/m^{2}$ after relaxation, which is considered a high value. On the other hand, the reconstructed models are most stable, having surface energies near of $620 \ \ mJ/m^{2}$ and $926 \ \ mJ/m2$ for Zn-terminations (\cref{fig:models}-d) and S-terminations, respectively. A Zn-terminated $ (2 \times 2) $ slab allowing to orient pairs of sulfur ions along the [110] direction are been simulated, with a similar surface of $ 636 \ mJ/m^{2}$ (see supp. material Fig S4b)
Furthermore, the reconstructed (001) surface has presented a strong relaxation of the ions at the surface. Zn specimens moved toward the surface ($\Delta z = 1$\AA \ ), at almost the height level of the S ions, which move outward the surface as depicted in \cref{fig:models}-(d). This displacement confers to the surface an apparent excess of S ions and confirms the theoretical hypothesis suggested from the discussed XPS results depicted in \cref{xps} (d)-(e). 

In addition, such an excess of sulfur species is in good agreement with the experimental findings from XPS measurements, which has revealed a prevalence of sulfur species over Zn one. More information of such results can be achieved from SCLS calculations, as it can be related to quantities like segregation  \cite{ruban1999surface} . The calculated SCLS of S 2p peaks is $0.69 \ eV$,  $0.12 \ eV$ higher than the unreconstructed system, a shift that belongs to $S^{-2}$ state reported by experimental works on ZnS \cite{Min2012minerals}. Thus, an asymmetric broadening of the peak is expected for this species. This finding endorses the sulfur-segregation that has been suggested by the increase of the FWHM measurement post a high surface treatment, reported in \cref{xps}-d. Moreover, the S-rich-reconstructed ZnS surface scenario is favorable to promote new surface states into the forbidden zone. These states are responsible to tailing the gap, as reported from DOS calculations (supp. material), hence explaining the better semiconducting behavior of this structure compared to the unreconstructed one.



\section{Conclusion}

In this research ZnS single crystal surface was investigated by XPS, LEED, and DFT. From the interplay between experiment and theory, it is shown the formation of a stable S-rich-reconstructed surface that seems to be more conductive than conventional ZnS systems. The cation-defectiveness structure is addressed to the preparation conditions that promote the removal of Zn species, as noted through XPS experiments by the increasing of the $S / Zn$ ratio as a function of the annealing temperature. Notably, post the higher treatment, that ratio became even higher when compared as a function of the takeoff angle, suggesting S species segregating to the outermost layers of the ZnS surface. Further, the XPS peaks have appeared with the expected binding energies without external assistance like flood gun, highlighting the neutralization of the charging. In this scenario, the first LEED pattern for ZnS single crystal was achieved, revealing a $(1 \times 2)$ surface reconstruction. DFT calculations are corroborative to the experimental findings, revealing a stable behavior of the reconstructed system whose topmost layers are indeed S-richer and Zn deficient. Moreover, these theoretical results pointed out a decrease in the ZnS band gap due to surface states induced into the forbidden zone, explaining the better conductivity of the S-rich-reconstructed ZnS surface. Our research shows the driven-force to the formation of a cation-deficient ZnS surface and a pathway to obtaining it, as well as the impacts of that configuration on the electronic and crystallographic properties of the ZnS, which may open the way to new investigations exploring this scenario for applications in physical-chemistry and geology.

\section*{Acknowledgments}

PO and FS are gratefully for the financial support of the CNPq, FAPERJ, Alexander von Humboldt
Foundation and the Max-Planck Partnergroup Program grants. CA thanks the Fundação de Amparo à Pesquisa do Estado de São Paulo (FAPESP) for the grant 2022/12650-9. The authors thank the Surface and Nanostructures Laboratory at CBPF.






\newpage
\bibliography{Reference.bib}






\end{document}